# Comparative Analysis of Hyperspectral Image Reconstruction Using Deep Learning for Agricultural and Biological Applications

Md. Toukir Ahmed[1], Arthur Villordon[2], Mohammed Kamruzzaman[1]*

[1]Department of Agricultural and Biological Engineering, University of Illinois at Urbana-Champaign, Urbana, IL 61801, USA.

[2]LSU AgCenter Sweet Potato Research Station, Louisiana State University, Baton Rouge, LA 70803, USA

## Abstract

Hyperspectral imaging (HSI) has become a key technology for non-invasive quality evaluation in various fields, offering detailed insights through spatial and spectral data. Despite its efficacy, the complexity and high cost of HSI systems have hindered their widespread adoption. This study addressed these challenges by exploring deep learning-based hyperspectral image reconstruction from RGB (Red, Green, Blue) images, particularly for agricultural products. Specifically, different hyperspectral reconstruction algorithms, such as Hyperspectral Convolutional Neural Network - Dense (HSCNN-D), High-Resolution Network (HRNET), and Multi-Scale Transformer Plus Plus (MST++), were compared to assess the dry matter content of sweet potatoes. Among the tested reconstruction methods, HRNET demonstrated superior performance, achieving the lowest mean relative absolute error (MRAE) of 0.07, root mean square error (RMSE) of 0.03, and the highest peak signal-to-noise ratio (PSNR) of 32.28 decibels (dB). Some key features were selected using the genetic algorithm (GA), and their importance was interpreted using explainable artificial intelligence (XAI). Partial least squares regression (PLSR) models were developed using the RGB, reconstructed, and ground truth (GT) data. The visual and spectra quality of these reconstructed methods was compared with GT data, and predicted maps were generated. The results revealed the prospect of deep learning-based hyperspectral image reconstruction as a cost-effective and efficient quality assessment tool for agricultural and biological applications.

---

* *Corresponding author. E-mail:* mkamruz1@illinois.edu, *Tel:* + 1 217-265-0423, *Website:* https://abe.illinois.edu/directory/mkamruz1

1. Introduction

Hyperspectral imaging (HSI) has emerged as a powerful tool for non-invasive quality assessment of various products [1]. Unlike conventional RGB imaging, HSI captures a detailed spectral signature at each spatial location across hundreds of narrow bands [2]. The integration of this abundant spatial and spectral data forms hypercubes. Efficient analysis of these hypercubes can provide relevant qualitative and quantitative information. This information can be crucial in identifying the chemical composition, material properties, and other physical characteristics of objects within an image [3]. Furthermore, HSI facilitates the visualization of the spatial distribution of quality attributes, providing a powerful tool for real-time monitoring and control in various applications. The wide range of applications, including but not limited to agriculture, medical imaging, remote sensing, and biology, highlight HSI's adaptability and extensive implications [4–8].

Despite its significant potential, HSI technology faces several challenges that hinder its widespread adoption [9–11]. The vast volume of spectral data from HSI systems comes at the cost of increased complexity, often rendering these systems impractical for real-time applications [11].

Conventional HSI systems, which rely on spatial or spectral scanning methods (e.g., push-broom or variable-filter systems), are typically characterized by bulky designs, moving parts, and extended exposure times [9]. These features make them less suitable for dynamic scenarios and contribute to their high cost, limiting their accessibility. While recent advancements in snapshot HSI technology have made strides towards real-time spectral image acquisition, they often compromise on spatial and spectral resolution [11]. Thus, a balance between performance and practicality in HSI systems remains a critical area of ongoing research and development.

In response to the limitations of traditional HSI systems, there has been a significant increase in interest in developing methods for reconstructing hyperspectral data from standard RGB images [11]. This approach presents a less costly and more widely accessible alternative. However, the task is inherently challenging because of the substantial reduction in detail when compressing the extensive spectral data of HSI into the constrained spectrum of RGB imaging [12]. Recent breakthroughs in deep learning offer novel opportunities for addressing this challenge [13]. Deep learning-based algorithms have shown remarkable potential in accurately reconstructing hyperspectral images from RGB data, even in scenarios with low-quality input images [11]. The reconstruction algorithms can incorporate good spatial resolution and provide spectral information [2]. These advancements significantly enhance the accessibility and usability of hyperspectral data in various fields, leading to more precise and well-informed decisions. This is especially beneficial in agriculture, where hyperspectral data increasingly replaces traditional methods based on standard RGB data [14]. Notably, initiatives like the New Trends in Images Restoration and Enhancement (NTIRE) spectral reconstruction challenges [9] have significantly contributed to promoting the innovation of reconstruction algorithms. However, the complete integration of these sophisticated algorithms, particularly in evaluating the quality of agricultural and biological products, has not been fully achieved.

The application of HSI reconstruction remains relatively underexplored within the specific context of agricultural and biological products. Only a limited number of studies have explored this field [15,16]. For instance, J. Zhao et al. [15] applied the HSCNN-R algorithm to assess the soluble solid content (SSC) in tomato samples, getting an $R^2$ of 0.51. More recently, Yang et al. [16] applied the pre-trained MST++ method to develop inversion models for assessing the psychological parameters of rice, with an $R^2$ of 0.40 for predicting the SPAD value. However, these studies represent only a fraction of the potential applications in the agricultural and biological domain, indicating a significant opportunity for further exploration and innovation. Many deep-learning methods have recently been introduced to reconstruct hyperspectral images from RGB [11]. This data-driven approach has many advantages over traditional model-based techniques. The conventional approaches often depend on too many tuning parameters, and when the environment or the dataset changes, the model parameters need to retune again. The data-driven deep learning approaches do not have this shortfall, as only neural network training is required when the dataset changes. Therefore, this study seeks to fill this gap by pioneering the application and comparative analysis of cutting-edge deep learning-based hyperspectral reconstruction algorithms, namely HSCNN-D, HRNET, and MST++, specifically for predicting the dry matter content in sweet potatoes (*Ipomoea batatas* L.). The specific objectives of this study were to (a) analyze spectral data from hyperspectral images of sweet potatoes, (b) select and explain the importance of feature wavelengths, (c) reconstruct hyperspectral images using three reconstruction algorithms and rigorously assess the visual and spectral quality of these reconstructions, and (d) predict and visualize the spatial distribution of dry matter content in sweet potatoes using both ground truth and reconstructed images.

2. Materials and methods

2.1 Sample collection and preparation

Sweet potato samples from three distinct cultivars, namely "Bayou Belle," "Murasaki," and "Orleans" were collected from the Sweet Potato Research Station at Louisiana State University AgCenter, Louisiana. A total of 141 sweet potatoes were investigated, including Bayou Belle (50), Murasaki (40), and Orleans (51). Following their collection, all samples were subsequently stored at a consistent room temperature of 25°C for 24 hours before image acquisition and reference measurements.

2.2 Image acquisition and correction

Before conducting the reference analysis, hyperspectral images of sweet potato samples were captured using a Specim V10E IQ line-scan hyperspectral camera designed for visible near-infrared (VNIR) exploration (Specim, Spectral Imaging Ltd., Oulu, Finland). The camera assembly consisted of a lens, imaging spectrograph, and CMOS sensor. Operated with Specim IQ Studio software, the handheld Specim IQ camera featured a 4.3-inch touch screen and 13 buttons for control. The camera featured 512 spatial pixels and 204 spectral bands, with a 7 nm spectral resolution spanning 400-1000 nm. The configuration consistently produced 2-D square images with a resolution of 512×512 pixels. These images were used to construct a 3-D hypercube with dimensions of 512×512×204, which contained a large dataset of over 53 million data points per sample. Each sample was imaged twice from two perspectives: one from a top-down view and one from underneath after being flipped. Symmetrically positioned two 750 W tungsten halogen lamps (ARRILITE 750 Plus, ARRI, Germany) ensured even field-of-view illumination. Hyperspectral images of a 99% reflective white surface ($I_W$) and reference images of a dark background ($I_D$ ~0% reflectance) were acquired to address non-uniformity and noise. Subsequently, Eq. 1 was applied to correct the unprocessed hyperspectral images ($I_{R0}$), which were then saved in band-interleaved-by-line (BIL) format for further analysis.

$$I_R = \frac{I_{R0} - I_D}{I_W - I_D} \quad (1)$$

2.3 Reference data collection

After acquiring the spectral data, the samples underwent immediate processing to assess their dry matter content (DMC). The commonly employed method for determining DMC in agricultural products involves utilizing an oven-drying technique [17]. Firstly, the weight of each sample was measured. Then, they were dried for 24 hours at a temperature of 103°C until a consistent mass was achieved, following the ASABE standard S358.1–2012 Moisture Measurement-Forage guidelines [17]. This process allowed for accurate determination of the dry weight. Finally, three replicates were conducted, and the results were averaged to calculate the dry matter content for each sample using Eq. 2.

$$DMC\% = \frac{W_d}{W_t} \times 100 \quad (2)$$

Here, DMC% is the dry matter (%) for a sample, $W_d$ is the dry weight of a sample in grams, and $W_t = W_d + W_m$, is the total weight of a sample in grams containing both the dry weight of the sample ($W_d$) and the weight of moisture content present in the sample ($W_m$).

2.4 Image segmentation and extraction of spectral data

Image segmentation is an essential initial step in extracting spectral information from hyperspectral images [18]. It involves identifying regions of interest (ROI) within the tested sample by creating a mask that distinguishes them from the background. In this study, the mask was generated by using the differences in reflectance between two specific bands (602 nm and 452 nm) to separate the sweet potato samples from the image's background (Supplementary information, Fig. S1). This mask subsequently served as the ROI for extracting reflectance spectral information from the hyperspectral images. A single mean spectrum was derived for each hyperspectral image by calculating the mean of all pixel spectra within each ROI to derive the reflectance spectra from the ROIs. Ultimately, the spectra for a sample were obtained by calculating the average of the two spectra derived from the corresponding images of that sample, given that each sample was imaged twice.

2.5 Features selection from the extracted spectra

Spectral signatures often contain redundant and unnecessary information across the entire spectral range, negatively impacting training precision [2]. To enhance the predictive capabilities of multivariable analysis and simplify the model, researchers have focused on selecting informative features from the numerous closely spaced spectral attributes [19]. Consequently, significant efforts have been directed toward devising and evaluating effective feature selection techniques [20]. Nevertheless, due to each experimental dataset's unique and intricate nature, no universally optimal method can be applied to all datasets [21]. Evolution-based optimizing techniques for analyzing multicollinearity data have recently gained popularity [19,22]. In this study, GA was employed to select key feature wavelengths for predicting the DMC of sweet potatoes after spectral data extraction. Drawing inspiration from the mechanisms of natural evolution, GA efficiently identifies the best combinations of features by employing a series of operations: selection, crossover, and mutation [23,24]. These genetic operators were carefully adjusted to optimize feature selection, including the selection method: "Tournament," crossover method: "Single-Point," mutation rate: 3 %, and elitism rate: 50%. The selected wavelengths were later used to construct a data hypercube for reconstructing hyperspectral images.

2.6 Hyperspectral image reconstruction

2.6.1 Ground-truth (GT) hypercube construction

The ground-truth (GT) hypercubes were exclusively built using the λ key wavelengths chosen by the GA. Consequently, each hypercube possessed dimensions of 512×512× λ. Given that each sample underwent imaging twice from different viewing angles, a total of 282 of these hypercubes were generated for the 141 samples. These hypercubes were later used as label data for the hyperspectral image reconstruction algorithms.

2.6.2 Input RGB image preparation

Specim IQ VNIR camera rendered a set of RGB images (282 in total) from hyperspectral images according to CIE Standard Illuminant D65 with the CIE 1931 2° Standard Observer using gamma correction ($ɣ = 1.4$). This camera combines wavelengths of 599 nm, 549 nm, and 449 nm to simultaneously create an RGB image in PNG (Portable Network Graphics) format while capturing HSI images. However, the reconstruction algorithms exclusively accepted RGB inputs in the JPEG (Joint Photographic Experts Group) format. Thus, the PNG files were converted to JPEG format. Additionally, the ROIs established in section 2.4 were consistently applied to these converted images. Ultimately, these segmented RGB files served as the input data for the reconstruction algorithms.

2.6.3 Hyperspectral image reconstruction algorithms

**2.6.3.1 HSCNN-D**

HSCNN-D is an advanced CNN-based hyperspectral image reconstruction method [25]. It utilizes densely connected layers, as illustrated in Fig. 1a, to effectively transform RGB images into hyperspectral data. In contrast to the deep residual network named HSCNN-R, HSCNN-D implements an innovative strategy wherein the residual block is substituted with a more complex structure, consequently deepening the network to achieve improved precision. HSCNN-D includes a unique path-widening fusion integrated within its dense architecture to increase the number of forward paths to enhance the size of the ensemble, thereby boosting the overall capacity and augmenting the model's overall performance. A key advantage of this dense structure is its ability to significantly reduce the issue of gradient vanishing during the training process. HSCNN-D achieves this by enabling the $k^{th}$ layer of the network to receive inputs from all preceding layers (as detailed in $f_0, f_1, \ldots, f_{k-1}$ in Eq. 3). This approach effectively resolves the challenges associated with signal propagation in increasingly deep networks. In addition to facilitating quicker convergence, the model's compact design is particularly suited for hyperspectral reconstruction tasks. It allows for the input of an RGB image and outputs a hyperspectral image with multiple channels, adeptly handling the inherent reduction in channels. The primary difference between the input and output is attributed to the reduction in the number of channels. Implementing a concatenation operator in each dense block further increases the channel count, leading to a more precise reconstruction model. Notably, HSCNN-D won the NTIRE-2018 spectral reconstruction challenge for clean images, attesting to its superior performance.

$$f_k = c_k([f_0, f_1, \ldots, f_{k-1}]) \tag{3}$$

Where $c_k(.)$ refers to the $k^{th}$ convolution layer and $[f_0, f_1, \ldots, f_{k-1}]$ denotes the concatenation of the features generated by preceding layers.

**2.6.3.2 HRNET**

Hierarchical Regression Network for Spectral Reconstruction from RGB Images (HRNET) [26] employs a 4-level hierarchical network structure (depicted in Fig. 1b), featuring interconnected networks at each level. This hierarchical design is strategically chosen to enhance information amplification compared to traditional encoder-decoder models. Within each level, the network architecture includes convolution operations followed by residual dense blocks and residual global blocks that employ multilayer perceptron (MLP) for attention computation, thereby enriching contextual understanding. A series of operations are performed at each network tier, including inter-tier integration, artifact elimination, and global feature extraction. For inter-level learning, the output features obtained from lower-level networks are subjected to pixel shuffle, concatenation with the current level, and further convolutional layer processing to standardize the channel count. The PixelUnShuffle layers are employed to down-sample the input at each level without increasing the number of parameters. This approach keeps the total pixel count of the input constant while reducing its spatial resolution. In contrast, the learnable PixelShuffle layers are used to up-sample feature maps, simultaneously decreasing the number of channels to facilitate connections between levels. HRNET uses a residual dense block consisting of 5 densely connected convolutional layers and a residual connection to mitigate artifacts effectively. The residual global block, featuring a shortcut connection to the input, is employed to extract attention for distant pixels using MLP layers. Given that the bottom level contains the most compact features, a $1 \times 1$ convolutional layer is appended to the end of the bottom level to enhance tone mapping by adjusting channel weights. The two intermediate levels operate on features at different scales. At the same time, the top level utilizes the most blocks to comprehensively integrate features and effectively reduce artifacts, resulting in the generation of high-quality spectral images. Regarding training, HRNET employs only the $L_1$ loss (as defined in Eq. 4) to optimize PSNR. Notably, HRNET has achieved recognition as the winning method in the NTIRE-2020 spectral reconstruction challenge, specifically for real-world images.

$$L_1 = E||\text{F}(X) - y||_1 \tag{4}$$

Where X is input, Y is output, and $F(.)$ is the HRNET algorithm.

**2.6.3.3 MST++**

MST++ is a Multi-Stage Spectral-Wise Transformer for Efficient Spectral Reconstruction [27]. This transformer-based architecture (Fig. 1c) represents an advancement over preceding hyperspectral reconstruction models by effectively addressing two key challenges: capturing extensive long-range dependencies and leveraging self-similarity priors. MST++ introduces a novel solution to these challenges, offering the advantage of being considerably lighter in weight compared to competitive models. Notably, MST++ addresses common issues faced when applying transformers directly to image reconstruction. Firstly, hyperspectral image reconstruction is characterized by spatial sparsity and spectral richness, whereas transformers primarily excel at capturing interactions among spatial regions. Secondly, the computational complexity of multi-head self-attention—the foundational component of transformers—is a notable concern. To tackle these issues, MST++ introduces S-MSA (spectral-wise multi-head self-attention), treating spectral feature maps as tokens for calculating self-attention across the spectral domain rather than the spatial domain. These S-MSAs (Eq. 5) collectively form the self-attention blocks (SABs). The SABs are integral to constructing the Single-stage Spectral-wise Transformer (SST), designed in a U-shape to enable the extraction of spectral context at multiple resolutions. Ultimately, the MST++ architecture, composed of a series of SSTs, incrementally enhances the reconstruction quality, evolving from a coarse to a fine level. MST++ is the winning method of the NTIRE-2022 spectral reconstruction challenge.

$$S - MSA(X) = \left(Concat_{j=1}^{N}(head_j)\right)W + f_p(V) \tag{5}$$

Where $W \in R^{C \times C}$ are learnable parameters, $f_p(.)$ Denotes the function to produce position embedding and $V \in R^{HW \times C}$ is the value that is generated from the linear projection of $X \in R^{H \times W \times C}$.

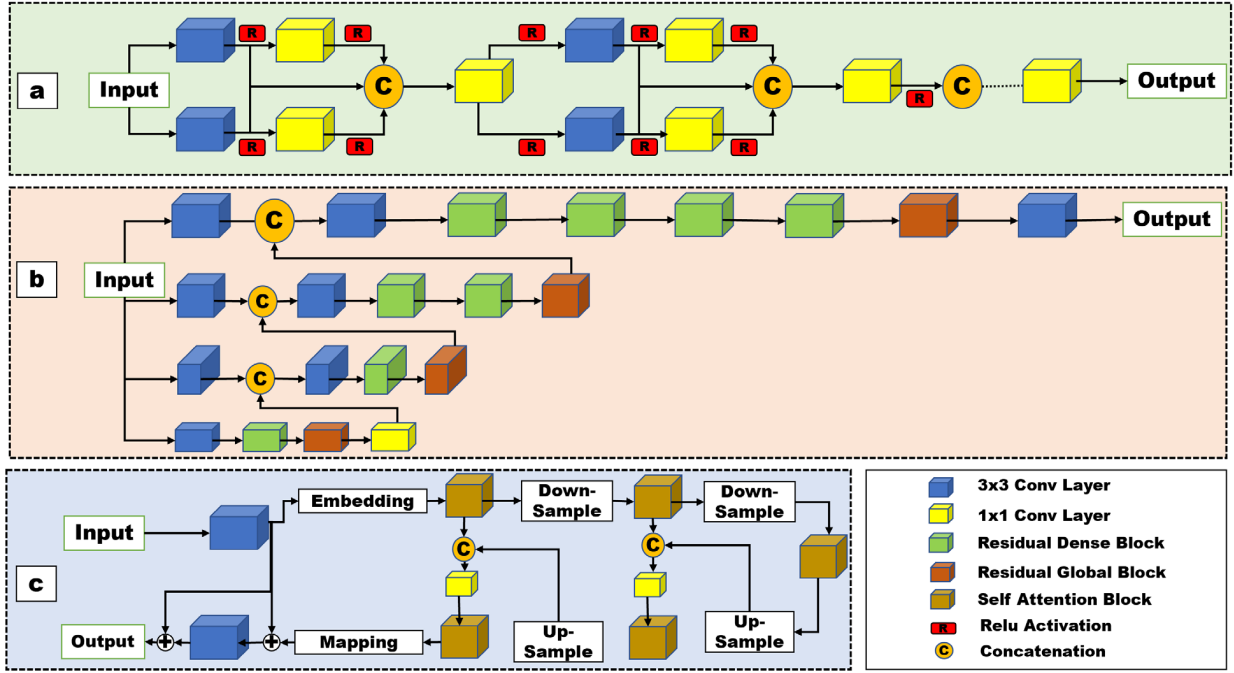

Fig. 1: Structure of the reconstruction algorithms (a) HSCNN-D, (b) HRNET, and (c) MST++

2.6.4 Reconstructed image evaluation

The efficacy of image reconstruction models was evaluated using the three evaluation metrics: mean relative absolute error (MRAE), root mean square error (RMSE), and peak signal-to-noise ratio (PSNR) [28,29] Eq. 6-8 symbolize the computations involved in carrying out the metrics. Each model was trained until there was no discernible decline in training MRAE loss. The performance of the training model was also assessed for the validation and test set.

$$MRAE = \frac{1}{n}\sum_{i=1}^{n}\left(\left|I_{rc}^{(i)} - I_{gt}^{(i)}\right|/I_{gt}^{(i)}\right) \qquad (6)$$

$$RMSE = \sqrt{\frac{1}{n}\sum_{i=1}^{n}\left(I_{rc}^{(i)} - I_{gt}^{(i)}\right)^2} \tag{7}$$

$$PSNR = 10 \times \log_{10}\left(\frac{255^2 \times n}{\left\|I_{rc} - I_{gt}\right\|^2}\right) \tag{8}$$

Where $I_{gt}$ and $I_{rc}$ represent ground truth and reconstructed hyperspectral images, respectively, and n denotes the total pixels.

2.6.5 Spectra extraction from the reconstructed images

The trained reconstruction models were used to reconstruct all images fed to the network. Each reconstructed image had the same dimension as the GT images (512×512× λ). The ROI masks computed in section 2.4 were again imposed on the reconstructed images, and the mean spectra were calculated. These spectra were finally analyzed for quality predictions of the samples.

2.7 Spectral analysis

2.7.1 Spectral model development

The spectral training model was constructed using the partial least squares regression (PLSR) technique, a widely adopted multivariate chemometric method for handling complex and correlated data [30]. PLSR identifies informative spectral regions for predicting reference values by mapping the spectral matrix (X) and reference variable (Y) into a subspace of latent variables (LVs) to maximize covariance. These LVs, unlike the original interconnected variables, are both informative and independent, leading to a simpler model. To ensure practicality and resistance to underfitting and overfitting, the optimal number of LVs was selected by the least root mean square error of cross-validation (RMSECV) using leave-one-out cross-validation (LOOCV) [30]. In addition to the ground truth and reconstructed spectra, a PLSR model was also developed using unprocessed RGB image data. Finally, the developed model was employed to predict the quality attribute of the samples.

2.7.2 Spectral model evaluation

Several statistical parameters, including coefficients of determination for calibration or training ($R_c^2$), root mean square error of training (RMSEC), coefficients of determination for validation ($R_v^2$), root mean square error of validation (RMSEV), coefficients of determination for prediction or testing ($R_p^2$), root mean square error of testing (RMSEP), and the ratio of prediction to deviation (RPD) were determined to evaluate the accuracy of the training models. A robust spectral model possesses high correlation coefficients ($R_c^2$, $R_v^2$, and $R_p^2$), a high RPD, and low errors (RMSEC, RMSEV, and RMSEP) [31].

2.7.3 Feature importance explanation

The importance of explainability in machine learning research has grown significantly in recent years, with explainable artificial intelligence (XAI) gaining widespread attention [32–34]. As AI advances rapidly, enhancing the transparency and interpretability of AI models has become increasingly complex [35,36]. In this context, the Shapley Additive exPlanation (SHAP) method has emerged as a valuable tool for XAI, aiding in unravelling the intricate relationships between multiple features and the predictions generated by computational models [35]. SHAP is a versatile, model-agnostic, game-theoretic approach that can provide feature attributions for individual instances [32]. In fields like chemometrics, SHAP values prove particularly beneficial as they offer a nuanced understanding of how various features, such as spectral wavelengths or chemical properties, influence a model's predictions. In this study, mean absolute SHAP values were computed to illustrate the importance of each feature in predicting outcomes for the PLSR model using the selected feature wavelengths.

2.7.4 Prediction map for visualizing response attribute

Hyperspectral imaging involves capturing a spectrum for each pixel in an image. A spatial prediction map can be generated by applying a calculated model to these pixels [37]. The study involved the development of PLSR models on the GT and reconstructed hyperspectral images to accurately predict the response attribute for individual pixels. The hyperspectral images were expanded into a two-dimensional column vector at specific wavelengths and then multiplied by the regression coefficient of the PLSR models. The matrix was transformed back into colored images with the same dimensions as the original images. Maps depicting the distribution of the response attribute were created using a color bar representing different colors to indicate spatial variations. Ultimately, a linear color scale was employed to depict the diverse distribution of the response attribute among the pixels in the distribution map.

2.8 Computational environment

The computations in this investigation were executed on the Google Colaboratory Pro (Colab Pro), a cloud computing platform (Google LLC, Mountain View, CA, USA). The Colab Pro virtual machine utilized a two-core Intel(R) Xeon(R) CPU @ 2.30GHz and 25 GB of RAM and an extended lifetime, and a Tesla P100 GPU (NVIDIA, Santa Clara, CA, USA) with 16 GB RAM. The analyses were conducted using Python 3.9. Open-source Python libraries [38], such as Scikit-learn, PyTorch, and OpenCV, were employed for training the reconstruction models.

3. Results and discussion

3.1 Spectral feature selection, model training, and explanation

HSI involves the acquisition of spectral signatures from each spatial location utilizing numerous narrow spectral bands. However, these collinear feature wavebands pose challenges regarding efficient management and processing. To address this issue and improve the understanding of the results without compromising their accuracy, it is essential to identify a subset of key feature wavelengths from the spectral data [19]. Nonetheless, due to each experimental dataset's inherent complexity and uniqueness, there is no universally dominant approach for identifying these crucial wavelengths [21]. This study addressed the challenge of identifying key wavelengths by utilizing the GA. GA enabled the identification of seven key feature wavelengths (406, 412, 478, 655, 709, 817, 832 nm) that significantly contributed to predicting the DMC of sweet potatoes. These selected wavelengths span both the visible spectrum (400-700 nm) and the short-wave near-infrared (SW-NIR) range (700-1000 nm).

Subsequently, a PLSR training model was developed, considering the full spectral range (400-1000 nm) and the selected feature wavelengths. To evaluate the model's performance, the spectral dataset was randomly partitioned into three subsets: train (60%), validation (20%), and test (20%). Random sampling, a widely accepted method in spectroscopy [39], was employed to ensure an unbiased selection of samples from the population. Specifically, 85 samples were used for training, while 28 samples each were dedicated to validation and testing. The response variable, DMC, exhibited mean values of 25.80 ± 6.54% for the train set, 29.11 ± 8.82% for the validation set, and 27.16 ± 7.01% for the test set. The number of LVs was determined by minimizing the RMSECV using the LOOCV technique. Table 1 provides a comprehensive overview of the PLSR model's performance across the entire spectral spectrum and with the selected features. Notably, the PLSR model utilizing the selected features outperformed the model based on the full spectra despite eliminating 96.57% of wavelengths. The model based on selected features exhibited higher values for $R^2$ and RPD and lower RMSE, highlighting the feature selection process's efficacy and the model's robustness. Consequently, these superiorly selected feature wavelengths were subsequently employed as GT data for the reconstruction algorithms.

Table 1: PLSR model on the full spectral range (400-1000 nm) and selected (GT) features.

| Spectral features | LV | Training | | Validation | | Testing | | RPD |
|---|---|---|---|---|---|---|---|---|
| | | $R_c^2$ | RMSEC (%) | $R_v^2$ | RMSEV (%) | $R_p^2$ | RMSEP (%) | |
| Full | 7 | 0.92 | 1.81 | 0.92 | 2.44 | 0.88 | 2.36 | 2.91 |
| Selected (GT) | 7 | 0.93 | 1.65 | 0.93 | 2.22 | 0.92 | 1.96 | 3.51 |

The assessment of the impact of GT features on predictions and their individual contributions was explained through a feature importance plot incorporating SHAP values [32]. In the context of PLSR modelling, this SHAP feature importance plot provided insight into the input features (X) exerting the most significant influence on the predicted values of response attributes (Y), surpassing the insights offered by traditional metrics. The feature importance plot for the PLSR model utilizing GT wavelengths as features is displayed in Fig. 2. The length of each bar in the plot represents the importance of the feature, determined by calculating the mean absolute SHAP values across the entire prediction dataset. These features are vertically organized, with the most influential ones listed at the top in descending order. A broader range of SHAP values indicates a stronger impact on the model's output. Fig. 2 underscores that feature wavelengths selected from the short-wave near-infrared (SW-NIR) regions, notably 817 nm, 832 nm, and 709 nm, exerted the most substantial influence, with 817 nm being the most impactful in predicting DMC. This alignment is expected, as the top-selected features primarily belong to the third overtone region of O-H bonds (Castro-Reigía et al., 2024), which is closely related to the calculation of DMC involving the removal of moisture from the samples (Eq. 2). Conversely, the selected wavelengths from the visible region had the least impact on DMC prediction. Therefore, the accurate reconstruction of the GT spectra, particularly in the SW-NIR regions, will likely yield improved predictive outcomes for the model.

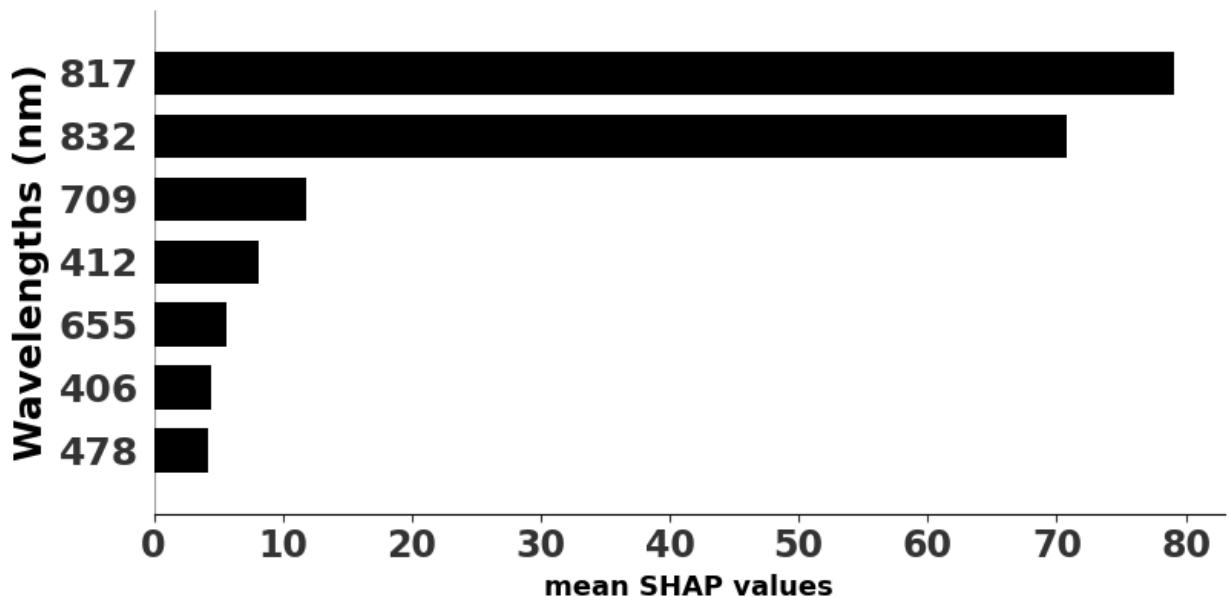

Fig. 2: Feature importance plot for explaining the model with GT spectra.

3.2 Training and validation of image reconstruction models

The initial step involved the extraction of the GT wavebands from each hyperspectral image, which was essential for data preparation in the development of image reconstruction models. These hyperspectral images, containing the GT bands, served as the label information, while the corresponding RGB images were used as inputs to the models. A total of 282 images, with two images per sample, were randomly distributed into three sets: train (60%), validation (20%), and test (20%). The models were trained using the Adam optimizer [41], with adjustments made to the momentum factors $\beta_1=0.9$ ($\beta_1=0.5$ for HRNET) and $\beta_2 = 0.99$. The batch size was set at 8, the patch size at 128, and the stride at 8. The initial learning rate for model training was set to $1e^{-4}$, with a gradual exponential decrease in learning rate during each epoch.

Following the initialization of all hyperparameters, each model underwent training for a total of 540 epochs, with each epoch consisting of 1000 iterations. The MRAE loss function was selected due to its consistent treatment of wavebands across different lighting conditions and its robustness against outliers [25]. Additionally, models trained with the MRAE loss function exhibited faster convergence in fewer epochs [15]. As illustrated in Fig. 3, the knee points for training loss [42] occurred at the $274^{th}$, $276^{th}$, and $286^{th}$ epochs, respectively, for HSCNN-D, HRNET, and MST++. After the knee-bending point, the loss remained constant. Consequently, the models trained at these highlighted epochs were selected for evaluation on the validation and test sets. The evaluation results of the trained hyperspectral reconstruction models are presented in Table 2.

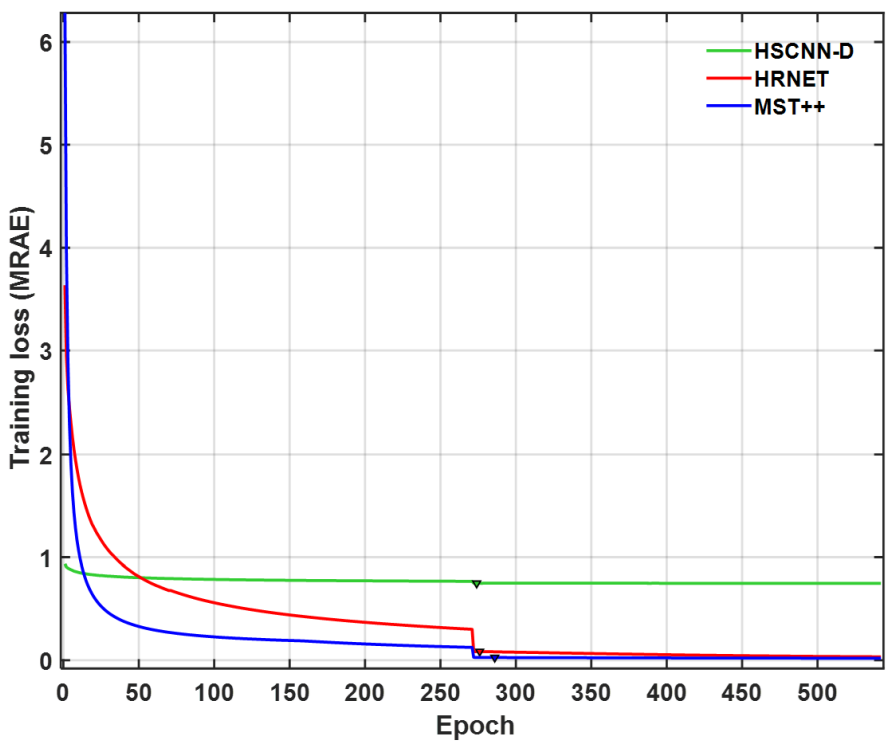

Fig. 3: Training MRAE loss of image reconstruction models

Table 2: Evaluation metrics of the image reconstruction models on validation and test set. Bold indicates the best performance in respective metrics.

| Method | MRAE$_{val}$ | RMSE$_{val}$ | MRAE$_{test}$ | RMSE$_{test}$ | PSNR (dB) | Execution time (hr) |
|---|---|---|---|---|---|---|
| HSCNN-D | 0.44 | 0.03 | 0.45 | 0.04 | 28.58 | 37.4 |
| HRNET | **0.06** | **0.02** | **0.07** | **0.03** | **32.28** | 49.5 |
| MST++ | 0.26 | 0.24 | 0.27 | 0.24 | 12.25 | **14** |

HRNET showed exceptional performance among the reconstruction methods, exhibiting the lowest MRAE (0.07) and RMSE (0.03) values and the highest PSNR scores (32.28 dB). Nevertheless, HRNET was found to be inefficient regarding time (49.5 hrs) and space complexity (3.17 e$^7$), as illustrated in Fig. 4. Conversely, MST++ emerged as the most efficient model in computational (0.08 G) and memory utilization (9.21 e$^4$). However, it performed poorly in terms of PSNR (12.25 dB) for the reconstructed images. The HSCNN-D method showcased superior PSNR performance (28.58 dB) compared to MST++ despite having the highest computational complexity (17.62 G), as highlighted in Fig. 4.

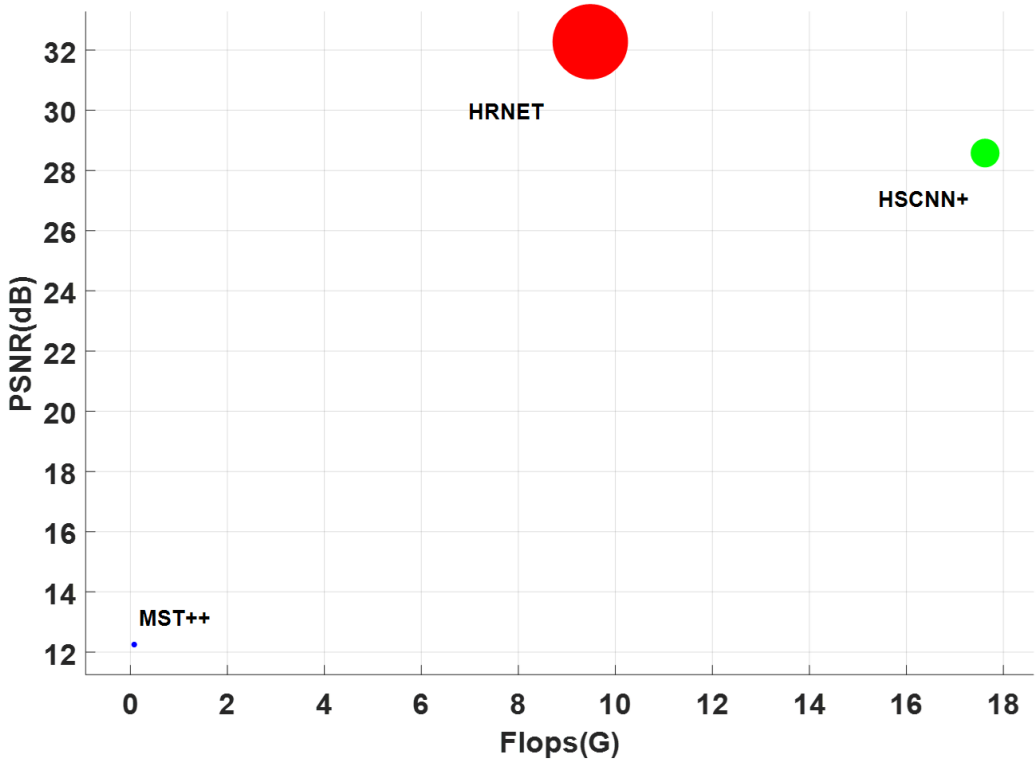

Fig. 4: PSNR- FLOPS comparisons among the reconstruction algorithms. The vertical axis is PSNR (performance), the horizontal axis is FLOPS (computational complexity), and the circle radius is memory cost.

3.3 Visual and spectral quality assessment

Following the successful execution of the model, a comparative analysis was conducted to assess the visual and spectral quality of the reconstructed hyperspectral images compared to the GT images. Fig. 5 depicts both the GT and reconstructed images at selected spectral wavelengths. Upon visually examining Fig. 5, it becomes evident that the reconstructed images closely resemble the GT images. Specifically, the images generated by the HRNET method exhibited a remarkable visual similarity to the GT images. However, subtle discrepancies were observed in the pixel intensities of the reconstructed images, particularly in the 655 and 709 nm wavebands. Conversely, the HSCNN-D method achieved a decent reconstruction of the images. Nevertheless, significant discrepancies in pixel brightness were observed in the lower-right corner of the reconstructed images compared to the GT images for the image at 832 nm, illustrated in Fig. 5. In contrast, the MST++ method displayed a significant deviation from the GT images in terms of pixel intensities. This method struggled to accurately reconstruct some pixels along the sample's edges, as indicated in Fig. 5.

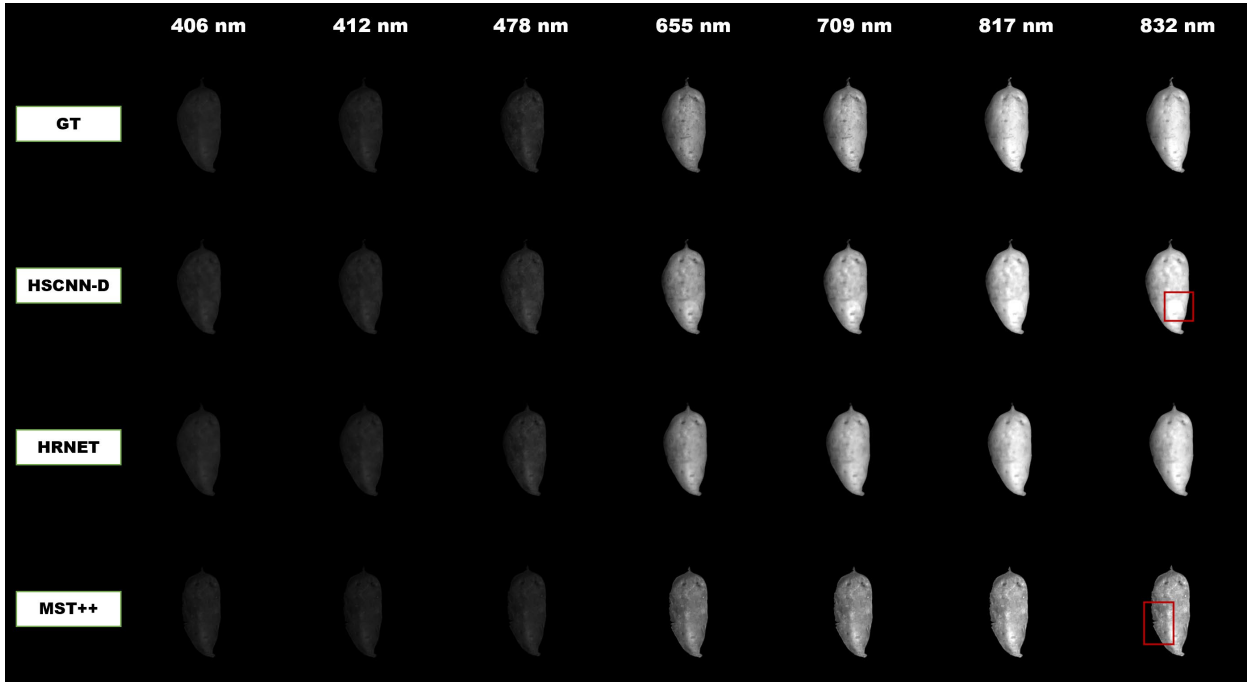

Fig. 5: GT and reconstructed images in the selected wavebands

After the image reconstruction, spectra were extracted from the reconstructed images, and the spectral quality of these reconstructions was evaluated. Fig. 6 displays the GT and reconstructed spectra. The spectra obtained from the reconstructed images by the HRNET method, as shown in Fig. 6, exhibited a remarkable resemblance to the GT spectra. This outcome was anticipated, given HRNET's exceptional reconstruction capability. HRNET exhibited a substantial similarity to the GT spectra in both the visible and SW-NIR regions, with only minor variations observed, particularly in the 655 and 709 nm wavebands. Conversely, the reconstructed spectral magnitude values derived from the HSCNN-D method closely corresponded to those of the GT spectra for wavebands in the visible region. However, notable discrepancies emerged in wavebands within the SW-NIR region, as depicted in Fig. 6. Lastly, the reconstructed spectra from the MST++ method exhibited significant deviations from the GT spectra, consistent with its subpar reconstruction performance. The analysis of reconstruction performance reveals a distinct trade-off between the quality of reconstruction and the time and space complexity of the employed algorithms.

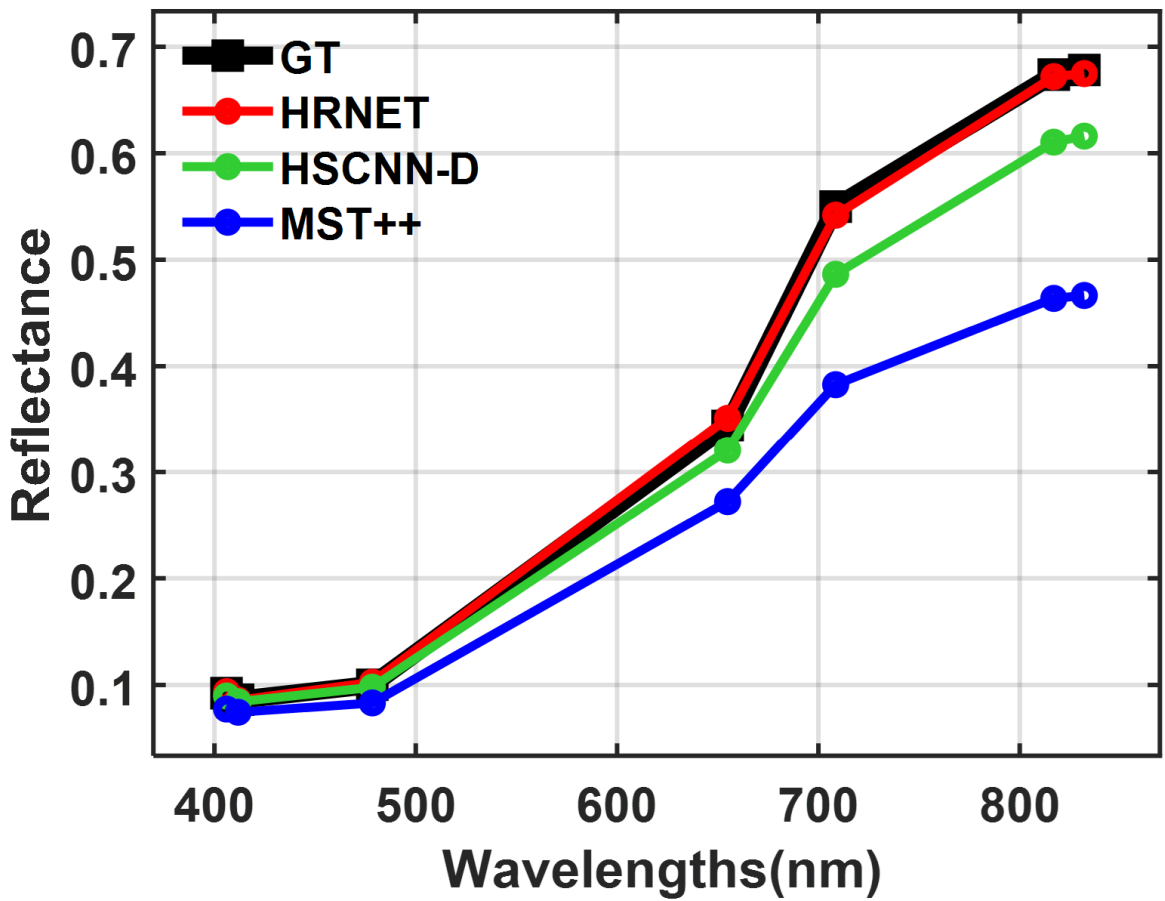

Fig. 6: Assessment of the GT and reconstructed spectra

3.4 Prediction of quality attribute using the reconstructed spectra

Following the spectra extraction from the reconstructed hyperspectral images, PLSR models were again developed using these reconstructed spectra. The efficacy of these PLSR models in predicting the DMC content of sweet potatoes is illustrated in Table 3. The performance of these models was profoundly affected by the quality of the spectral reconstruction methods implemented. Among these methods, the PLSR model applied to spectra reconstructed using the HRNET method performed the best, primarily due to its superior reconstruction performance. This model exhibited improved quality, as evidenced by higher $R^2$ (0.88-0.92) and RPD (2.93) values, lower RMSE values, and a smaller difference in RMSE values compared to the model applied on the full wavebands. This outcome aligns with expectations, given that HRNET efficiently reconstructed the spectra, especially in the SW-NIR region, which is crucial for DMC prediction (as depicted in Fig. 2). Conversely, the PLSR model utilizing the HSCNN-D method demonstrated decent performance but suffered from a slight deviation from the GT spectra (Fig. 6), particularly in the SW-NIR region, which negatively affected its overall performance. The MST++ method, on the other hand, performed poorly due to its inferior reconstruction quality, resulting in the most significant deviation from the GT spectra. In addition to the reconstructed models, a PLSR model was developed using the unprocessed RGB image data of the samples, as detailed in Table 3. Importantly, all reconstructed spectral models featured in the study showed improved performance over the RGB data-based model across all performance metrics. This enhanced performance of the reconstruction models clearly indicates that hyperspectral reconstructions offer significant advantages over traditional RGB imagery, particularly in improving the accuracy of quality estimation for sweet potatoes.

Notably, no study has yet reported the utilization of reconstructed spectra to predict the DMC of any agricultural or biological product. However, Zhao et al. [15] applied the HSCNN-R algorithm to reconstruct hyperspectral images and assess the SSC of tomatoes. They achieved an $R^2$ of 0.51 for predicting the SSC of tomatoes based on reconstructed spectra. Therefore, the exceptional performance observed in this study could potentially inspire the adoption of hyperspectral reconstruction for quality assessment across various domains.

Table 3: Performance of PLSR models on the reconstructed spectra and RGB data

| Method | LV | Training | | Validation | | Testing | | |
|---|---|---|---|---|---|---|---|---|
| | | $R_c^2$ | RMSEC (%) | $R_v^2$ | RMSEV (%) | $R_p^2$ | RMSEP (%) | RPD |
| HSCNN-D | 7 | 0.92 | 1.78 | 0.89 | 2.84 | 0.84 | 2.74 | 2.51 |
| HRNET | 7 | 0.92 | 1.81 | 0.89 | 2.85 | 0.88 | 2.35 | 2.93 |
| MST++ | 7 | 0.54 | 4.42 | 0.61 | 5.42 | 0.65 | 4.04 | 1.70 |
| RGB | 3 | 0.44 | 4.84 | 0.52 | 6.00 | 0.56 | 4.56 | 1.51 |

3.5 Visualization of the quality attribute

Hyperspectral image analysis offers a compelling advantage in understanding predicted quality attribute distributions in a spatial context compared to traditional spectroscopy [37]. These prediction maps demonstrate the spatial variability of quality attributes, showcasing the capability of hyperspectral imaging to provide detailed insights at the pixel level that go beyond what can be obtained from RGB images or human observation [18]. Regression models step in to interpolate the quality attribute distribution across the entire sample when precise concentration assessment in each pixel is unattainable through laboratory analytical methods. This study employed the PLSR model on the GT and reconstructed spectra using HRNET to visualize the respective hyperspectral images. The resulting prediction maps (Fig. 7) display the distribution of the DMC attribute across sweet potato samples. Notable differences emerge in the prediction maps between the GT and reconstructed images, primarily due to subtle variations in pixel intensities and deviations in predictive model performance. Overall, the prediction map on the reconstructed image demonstrates decent performance, particularly on the pixels located at the sample edges.

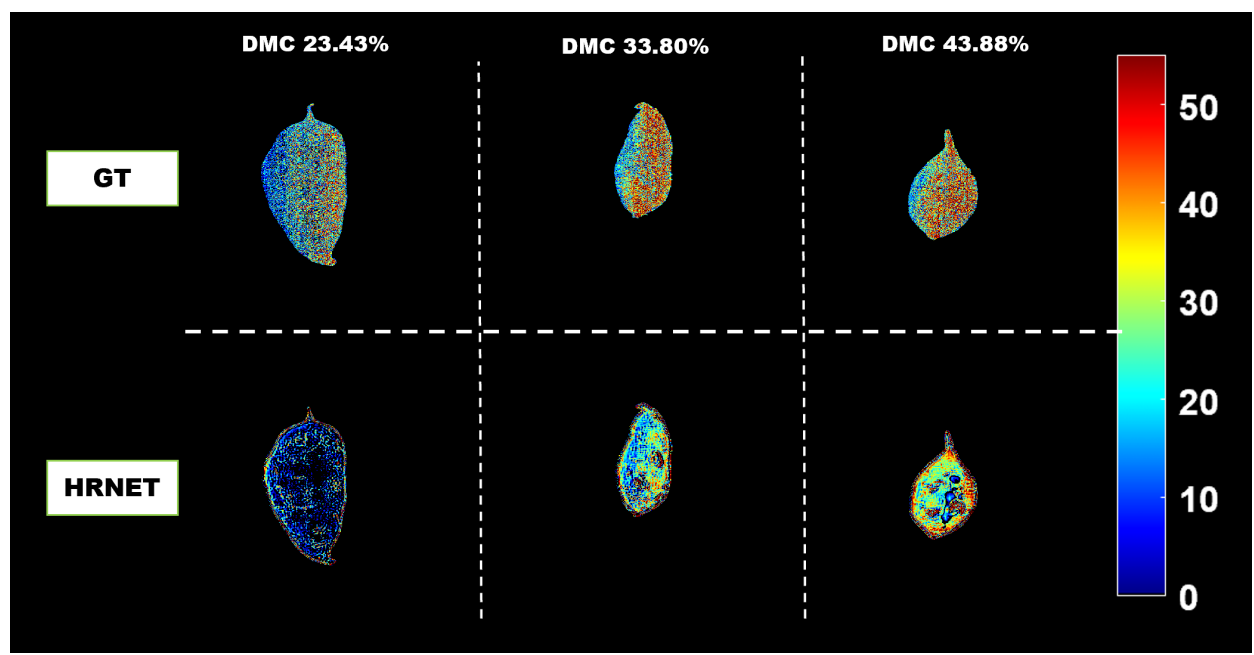

Fig. 7. Prediction map of DMC distribution in GT and reconstructed images (using HRNET); the unit of the color bars is the same as that of the corresponding attribute.

4. Conclusions

This study represents a significant advancement in applying hyperspectral image reconstruction for agricultural product quality assessment, particularly focusing on sweet potatoes. It demonstrates the efficacy of deep learning algorithms, including HSCNN-D, HRNET, and MST++, in transforming RGB images into hyperspectral data, overcoming the limitations of conventional HSI systems. The findings indicate that the reconstructed hyperspectral images and spectra closely resemble their ground-truth counterparts. This similarity is key in accurately predicting and visualizing the dry matter content of sweet potato samples. Consequently, this study fills a significant gap in the current HSI applications, presenting a cost-effective, efficient, and accessible approach for quality assessment in agricultural and biological products. Overall, this transformative approach will facilitate the implementation of hyperspectral imaging for end users using smartphones, digital cameras, and drones through cloud computing for various agricultural and biological uses. However, hyperspectral image reconstruction from RGB is an ill-posed problem and many combinations of reconstructed images are possible from the same RGB (known as the metamerism phenomenon). Therefore, it may introduce errors in reconstructed hyperspectral images due to the ill-posed nature of the inverse problem and the limitation of the RGB images. The reconstructed models may suffer from generalization issues as their performance can vary based on the domain, object, and characteristics of the imaging system. Moreover, effectively handling complex and dynamic agricultural settings such as varying illumination, occlusion, noise, and diverse backgrounds is challenging. Therefore, it is crucial not only to obtain application-specific reconstructed hyperspectral images from RGB images but also the careful selection and fine-tuning of appropriate deep-learning methods to potentially reduce the effects of metamerism and improve the accuracy and generalization of the reconstruction for real-time implementation and future advancements in hyperspectral imaging.

**Acknowledgment**

We greatly acknowledge Professor Arthur Villordon from the Sweet Potato Research Station at Louisiana State University AgCenter, Louisiana, for providing the sweet potato samples for this study.